\documentclass[pra,aps,showpacs,twocolumn,superscriptaddress]{revtex4}
\usepackage{graphicx,amsfonts,amsmath,amssymb,xcolor}
\begin{document}
\title{Refractory period of an excitable semiconductor laser with optical injection}
\author{B. Garbin\footnote{Corresponding author: bruno.garbin@inln.cnrs.fr}}
\affiliation{Universit$\acute{e}$ C\^{o}te d'Azur - CNRS, France}
\author{A. Dolcemascolo}
\affiliation{Universit$\acute{e}$ C\^{o}te d'Azur - CNRS, France}
\author{F. Prati}
\affiliation{Dipartimento di Scienza e Alta Tecnologia, Universit$\grave{a}$ dell'Insubria, via Valleggio 11, I-22100 Como, Italy}
\affiliation{CNISM, Research Unit of Como, via Valleggio 11, I-22100 Como, Italy}
\author{J. Javaloyes}
\affiliation{Departament de F\'isica, Universitat de les Illes Baleares, C/ Valldemossa km 7.5, 07122 Mallorca, Spain}
\author{G. Tissoni}
\author{S. Barland}
\affiliation{Universit$\acute{e}$ C\^{o}te d'Azur - CNRS, France}
\date{\today}
\begin{abstract}

Injection-locked semiconductor lasers can be brought to a neuron-like excitable regime when parameters are set close to the unlocking transition. Here we study experimentally the response of this system to repeated optical perturbations and observe the existence of a refractory period during which perturbations are not able to elicit an excitable response. The results are analyzed via simulations of a set of dynamical equations which reproduced adequately the experimental results.

\end{abstract}

\maketitle

\section{Introduction}
Excitability is a very general property of some nonlinear systems, which respond to external perturbations only if this perturbation overcomes a certain threshold. Beyond the original biological setting \cite{hodgkin,hodgkin2}, excitable systems have been studied for many years and in many different contexts such as chemistry \cite{che,nat}, optics and optoelectronics with laser with saturable absorber \cite{absorber,absorber2}, lasers with optical feedback \cite{giudici}, lasers with optical injection \cite{coullet,goulding,kelleher,barland} and resonant tunneling diodes \cite{RJI-OE-13}.

From the point of view of the nonlinear dynamics, several well defined scenario may lead to an excitable behavior, which may sometimes lead to closer analogies between a biological and a laser system than between two optical systems.
Beyond its academic interest, excitability has been intensively studied due to its large number of potential applications, among which we can cite neural architectures for computing and processing data \cite{maass} or, in optics, wavelength conversion and pulse reshaping due to the unicity of the response \cite{absorber}.

Several mathematical descriptions can provide a suitable phase space structure for a system to be excitable (see for instance \cite{izhikevich2006dynamical}). For the specific case of a laser locked to an external optical injection, it was theoretically shown that under some approximations (small detuning between slave and master laser and small injected power) the dynamics of the system close to the unlocking transition is slaved to that of the optical phase \cite{coullet}. Therefore its dynamics, described by Adler's equation \cite{adler}, is expected to be very similar to that of the $\theta$-neuron \cite{ermentrout1986parabolic,izhikevich2006dynamical}  which, as most excitable systems, is characterized by a refractory time \cite{izhikevich2006dynamical}. Indeed, the system may not respond to an external stimulation very shortly after responding to a first perturbation. Basically, the refractory time is the time it takes for the system to complete a full excitable orbit in phase space and to (exponentially slowly) come back to its stable fixed point.

Recently, the possibility to trigger an excitable response in a laser with optical injection applying a perturbation has been demonstrated for several kinds of perturbations: i) an incoherent pulse coming from a third laser \cite{garbin:14}, ii) a perturbation in the current of the injected laser, or iii) in the phase of the injecting laser \cite{control}. This last method has been shown to be the most efficient method. This is to be expected since the injected laser dynamics is slaved to that of the phase of the electric field and perturbing the phase of the driving beam gives rise to the stronger projection onto  the dynamics.

However, only the response to a single perturbation has been analyzed and, contrary to many other optical excitable systems where the refractory time has been directly or indirectly observed \cite{preacoplados,PhysRevA.85.031803,PhysRevA.65.033812,barbay,shastri2016spike}, the refractory period of this excitable system has not been measured yet.

In the following, we investigate the response to single or multiple perturbations of a laser with optical injection in the excitable regime. We observe that when the perturbation amplitude hardly overcomes the excitability threshold, not only the response is largely delayed, but the delay times show an increased dispersion, too. In addition, when we perturb the system twice with two perturbations close in time, the response to the second perturbation is largely influenced by the refractory period of the system, which again causes delayed and dispersed responses.

We analyze the experimental observations with the help of numerical simulations of a set of dynamical equations for an optically injected semiconductor laser. When the parameters are such that the relaxation oscillations are sufficiently damped, a situation similar to that of the experiment, the numerical simulations reproduce well the experimental data. Both the experiment and the numerical simulations suggest that, even if the pure phase model usually invoked to discuss excitability in this setting gives a very useful picture, it does not provide a complete description of the relaxation towards the original fixed point when the amplitude of the excitable spike is not arbitrarily close to zero and semiconductor medium dynamics may play a role in the definition of the refractory period.

\section{Experiment}
\begin{figure}[t]
\centering
\includegraphics[width=1.0 \columnwidth]{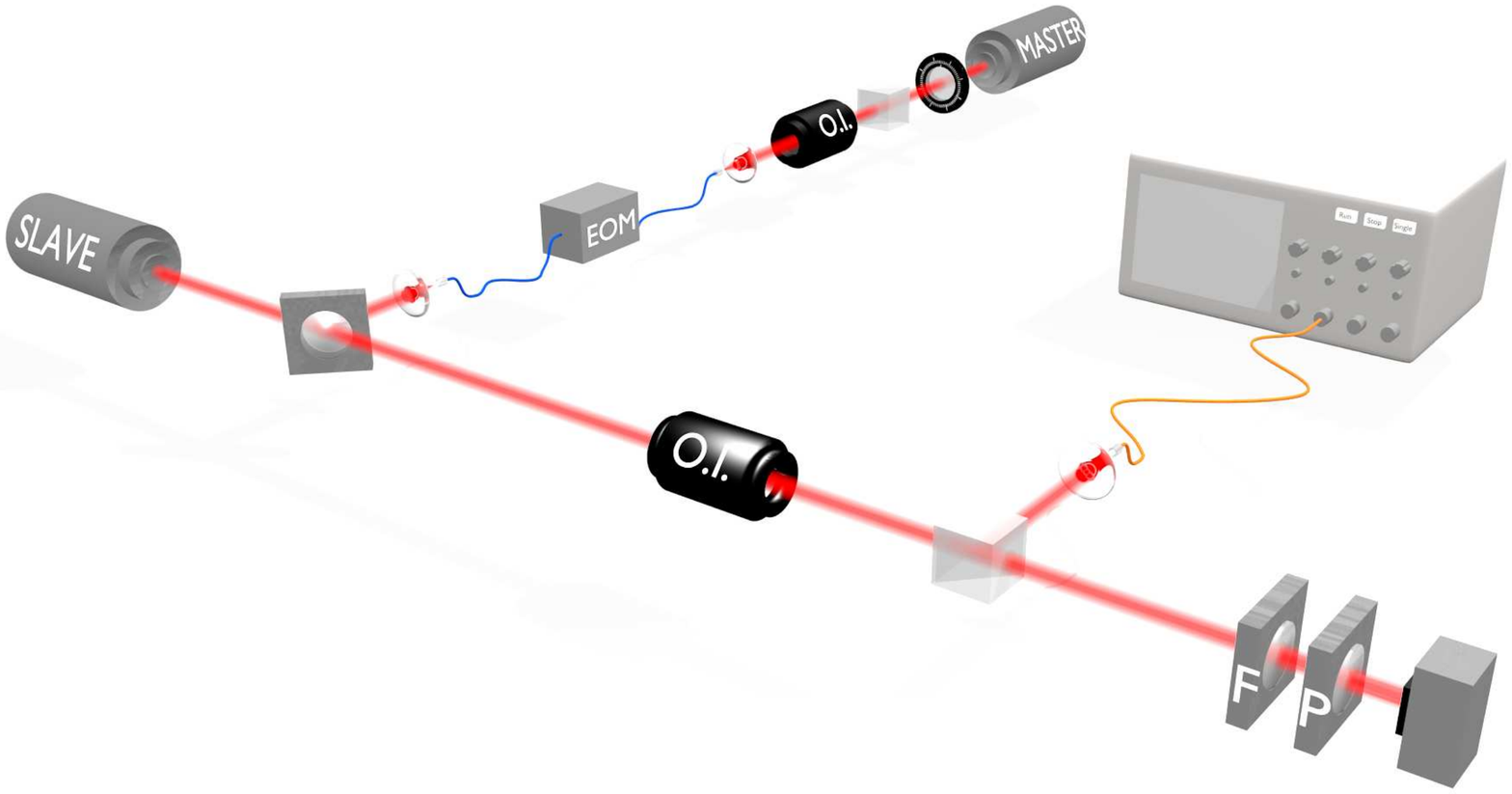}
\vglue-1.5cm
\caption{Schematic of the experimental set-up. MASTER: injection laser, SLAVE: injected laser, EOM: electro-optic modulator, O.I.: Optical isolator,
FP: Fabry-P\'erot interferometer.}\label{fig1}
\end{figure}

The experimental setup of the laser with optical injection, schematically shown in Fig. \ref{fig1}, is the same as in \cite{control}.

The master laser (injection laser) is an edge-emitter tuned by means of a grating. The intensity of the emitted beam can be controlled by a half-wave plate and a polariser placed on its path. An optical isolator with 30 dB isolation avoids bi-directional coupling. The master laser is coupled to a monomode polarisation maintaining fiber linked to an electro-optic modulator, which is used to apply a phase perturbation. The shape of the phase perturbation can be varied depending on what is needed (details below). Finally, the light coming form the master laser is injected into the slave laser through a 10\% reflection beam-splitter.

The slave laser is a single-longitudinal and -transverse mode VCSEL (ULM980-03-TN-S46), emitting on a single and linear polarization. Its threshold ($I_\mathrm{th}$) is about 0.2~mA. The output is directed to a half-wave plate and a polarizing beam-splitter, to be split and analysed by a scanning Fabry P\'erot interferometer (Finesse 110 and 71~GHz free spectral range) for one part. The remaining part is injected into an optical fiber, which guides light to a 9~GHz bandwidth photodetector. The signal is amplified by a RF amplifier (mini-circuit ZX60-14012L-S+) and then acquired by a real-time oscilloscope (DPO71254C 12.5~GHz, 100~GS/s).

The laser with optical injection has three control parameters: the detuning between the two lasers $\Delta$ defined as the frequency of the slave laser minus that of the master ($\Delta=\nu_S-\nu_M$), the injection strength of the master $P_\mathrm{inj}$, and the slave pumping current $I_\mathrm{sl}$.

When the detuning is weak enough the two lasers are locked in frequency with a constant relative phase. The size of the locking region as well as the kind of the unlocking bifurcation, in terms of detuning, depend on the injection strength  \cite{wieczorek_report}. The bifurcation which we are interested in here is the saddle-node on a circle, because of the presence of two fixed points (one stable, one unstable) when locked. At the  bifurcation the two merge, leaving a circle in the phase space, which corresponds to an oscillating system in time. This bifurcation appears for sufficiently low injected power. Preparing the system close to the bifurcation, when the two fixed points are close, we can observe an excitable behavior.

We place the system as close as possible to the unlocking boundary in order to minimize the excitability threshold.
The master laser frequency is kept constant and the detuning is varied by changing slightly the frequency of the slave laser by adjusting its pump current.
We will then use a phase modulation of the master beam as a perturbation.
%

The applied perturbation is stepwise, coming from a pulse generator (HP8133a) with 31.2 MHz repetition rate and 100~ps rise/fall time. Its output is amplified by the perturbation amplifier, a RF amplifier (photline DR-DG-10MO-NRZ), to get high perturbations (8~V max).
We pumped the slave laser at about $5I_\mathrm{th}$, the injection strength was about $2.5$ $\mu$W, and the detuning was $\Delta=-5.8$~GHz.

\begin{figure}[t]
\centering
\includegraphics[width=1.0 \columnwidth]{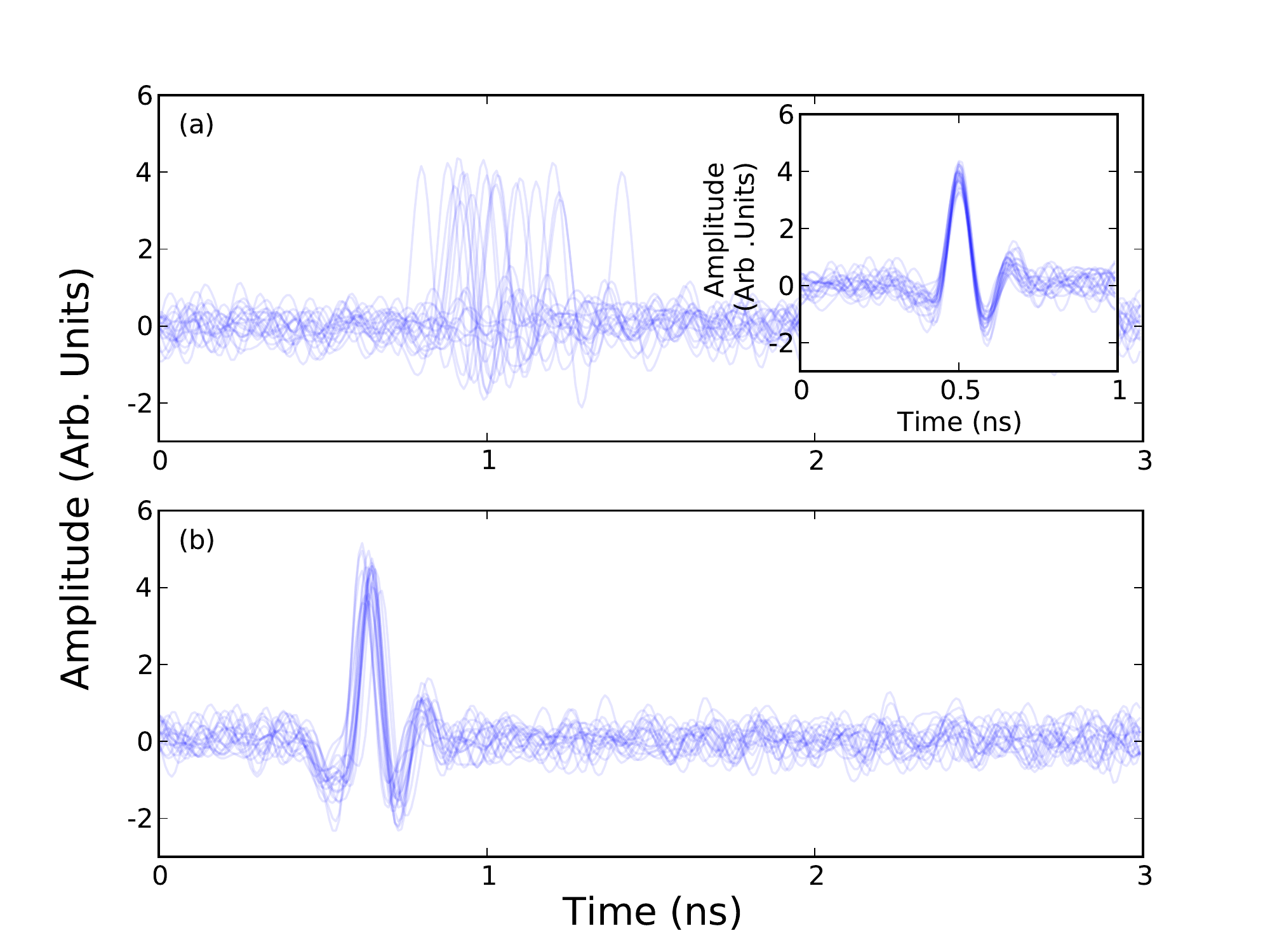}
\caption{Superposition of 20 excitable responses for different perturbation  with $I_\mathrm{sl} = 5I_\mathrm{th}$,
	$P_\mathrm{inj} = 2.5$ $\mu$W, and $\Delta = -5.8$~GHz.
(a): Perturbation amplitude of about 60$^\circ$, triggered on the perturbations.
Inset of (a): Same as (a), triggered on the maximum of the responses.
(b): Perturbation amplitude of about 160$^\circ$, triggered on the perturbations.}\label{fig2}
\end{figure}

Fig. \ref{fig2} shows typical examples of excitable responses. As we mentioned the system is essentially a phase oscillator when injected with a sufficiently low power. Then the intensity trace outgoing from the system shows a tiny excitable peak due to the phase-amplitude coupling. Indeed the excitable response corresponds to a $2\pi$ relative phase slip \cite{kelleher2}. The observed duration of pulses is about 100~ps. We observed down to 70~ps pulses which is the detection limit. The pulse duration depends on the exact condition of the detuning: the larger the detuning, the faster the pulses.
It must be noted that the perturbations giving rise to excitable responses are upward steps in the phase of the master laser beam, at least for the sign of $\Delta$ considered here.

Twenty excitable events are superimposed in the two panels of Fig. \ref{fig2}. These traces are triggered on the external perturbation which defines the origin of time. The upper and lower panel differ for the magnitude of the perturbation which is 60$^\circ$ in (a) and 160$^\circ$ in (b). In both cases, the pieces of time series are superimposed in such a way that the time at which the perturbation reaches the system is always the same, \textit{i.e.} the perturbations are synchronized. By comparing the two figures we observe that the distribution of the delay times is broader and its mean value is larger for the smaller perturbation. The inset of Fig. \ref{fig2}(a) represents the same responses as in Fig. \ref{fig2}(a), but this time the time traces are superimposed in such a way that the responses are synchronized. This allows to check that, in spite of the large dispersion in the firing time, the response of the system is indeed always the same.

These findings match very well the very first observations of excitable behavior \cite{hodgkin}, which showed that the amplitude of the perturbation, particularly close to the excitable threshold, is important for the latency before the creation of an excitable pulse. This is because when a system is close to an unstable fixed point, the time to escape from it depends logarithmically on the distance between the initial condition and the unstable fixed point itself.

\begin{figure}[t]
\centering
\includegraphics[width=1.0 \columnwidth]{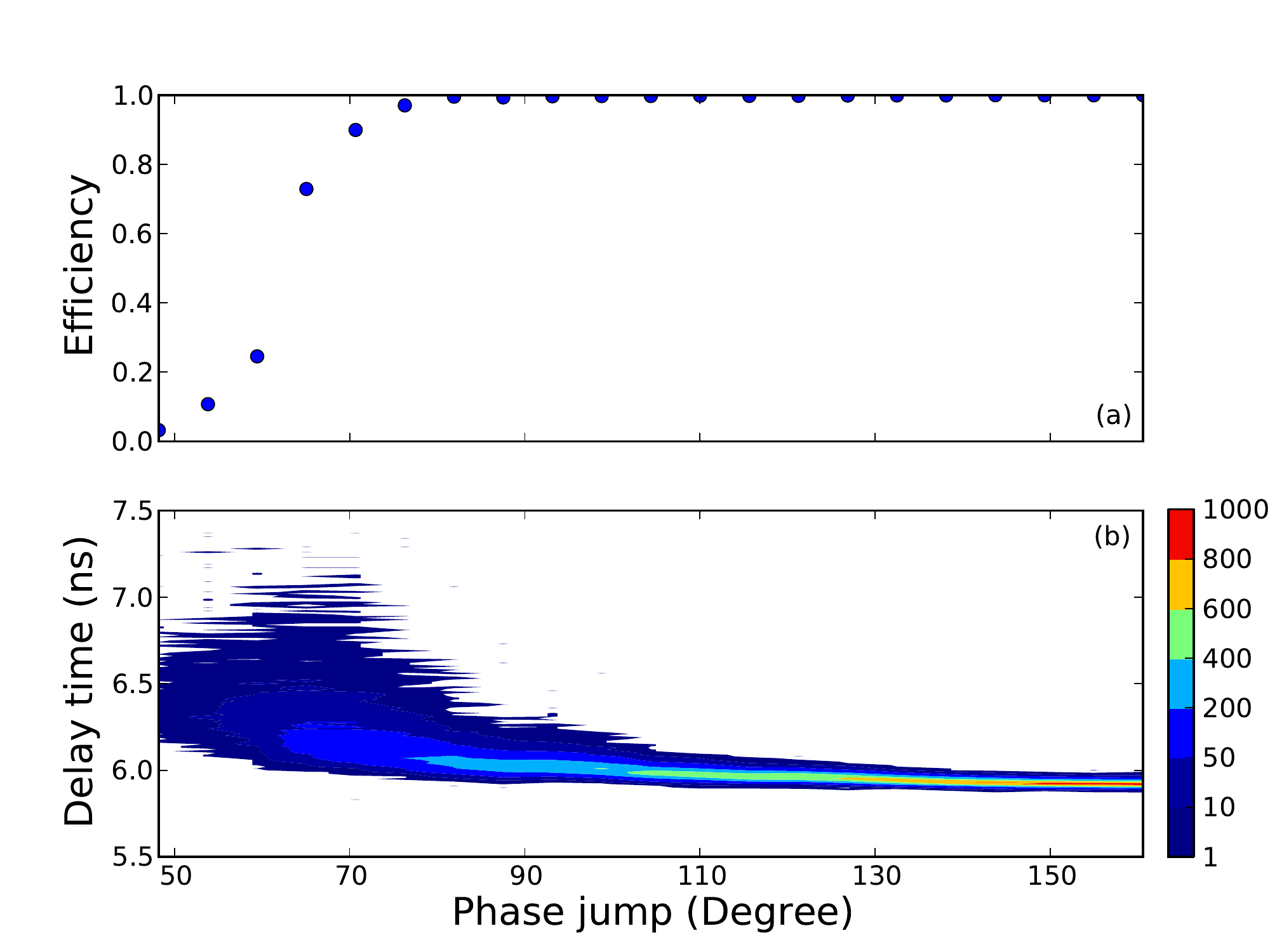}
\caption{Perturbation efficiency and response time to external perturbations. Same parameters as in Fig. \ref{fig2}. (a): Efficiency curve of the perturbation for more than 3000 events, defined as the number of excitable responses over the number of perturbations. (b): Associated evolution of the delay time histogram between the pulse and the trigger of the perturbation.}\label{fig3}
\end{figure}

The same information as in Fig. \ref{fig2} is contained in Fig. \ref{fig3} but for a much larger set of perturbation amplitudes. The efficiency curve, defined as the ratio of the number of excitable responses to the number of applied perturbations, is shown in Fig. \ref{fig3}(a). The existence of an excitability threshold (that we define here as the value of the perturbation for which the efficiency is 50\%), is clearly visible around 65$^\circ$; 100\% efficiency is reached at about 80$^\circ$. Some excitable responses of the third and the last point of Fig. \ref{fig3}(a) were respectively shown in panels (a) and (b) of Fig. \ref{fig2}.

One vertical cut of Fig. \ref{fig3}(b) represents a histogram of the delayed response time of pulses for one perturbation amplitude. Fig. \ref{fig3}(b) is constructed in such a way that we can see the evolution of this distribution with the perturbation amplitude.

For lower perturbations the excitable responses appear after a longer delay and the distribution of the response time is broader, with a width around 1.5~ns. Increasing the perturbation amplitude the distribution becomes more and more narrow and its mean decreases, as already noted in Fig. \ref{fig2}. For large perturbations the efficiency is about 100\% and the distribution is very sharp, with a width of about 0.1~ns for the largest amplitude. The mean delay time tends to about 5.9~ns. This value, however, is not the exact delay time of the excitable pulses with respect to the perturbation, because  it includes several other contributions: the time for the perturbation to actually reach the system (electrical signal propagation in cables and optical propagation from the modulator to the slave laser), the time for the optical excitable response to propagate towards the detector and the photodetector response time.

In order to measure the refractory period of two consecutive excitable pulses we perturbed the system in a slightly different way. The first perturbation is a short rectangular pulse, created by an electrical pulse generator (picosecond EPG-200B-0050-0250), with about 100~ps duration. The second is the same stepwise perturbation used before. The minimum technically attainable time interval between the two perturbations is 0.19~ns, measured at the middle of each rising edge. Due to the saturation of the amplifier placed between the electrical pulse generators and the optical phase modulator, we are not able to apply two identical steps and must apply first one pulse, followed by a step.
\begin{figure}[t]
\centering
\includegraphics[width=1.0 \columnwidth]{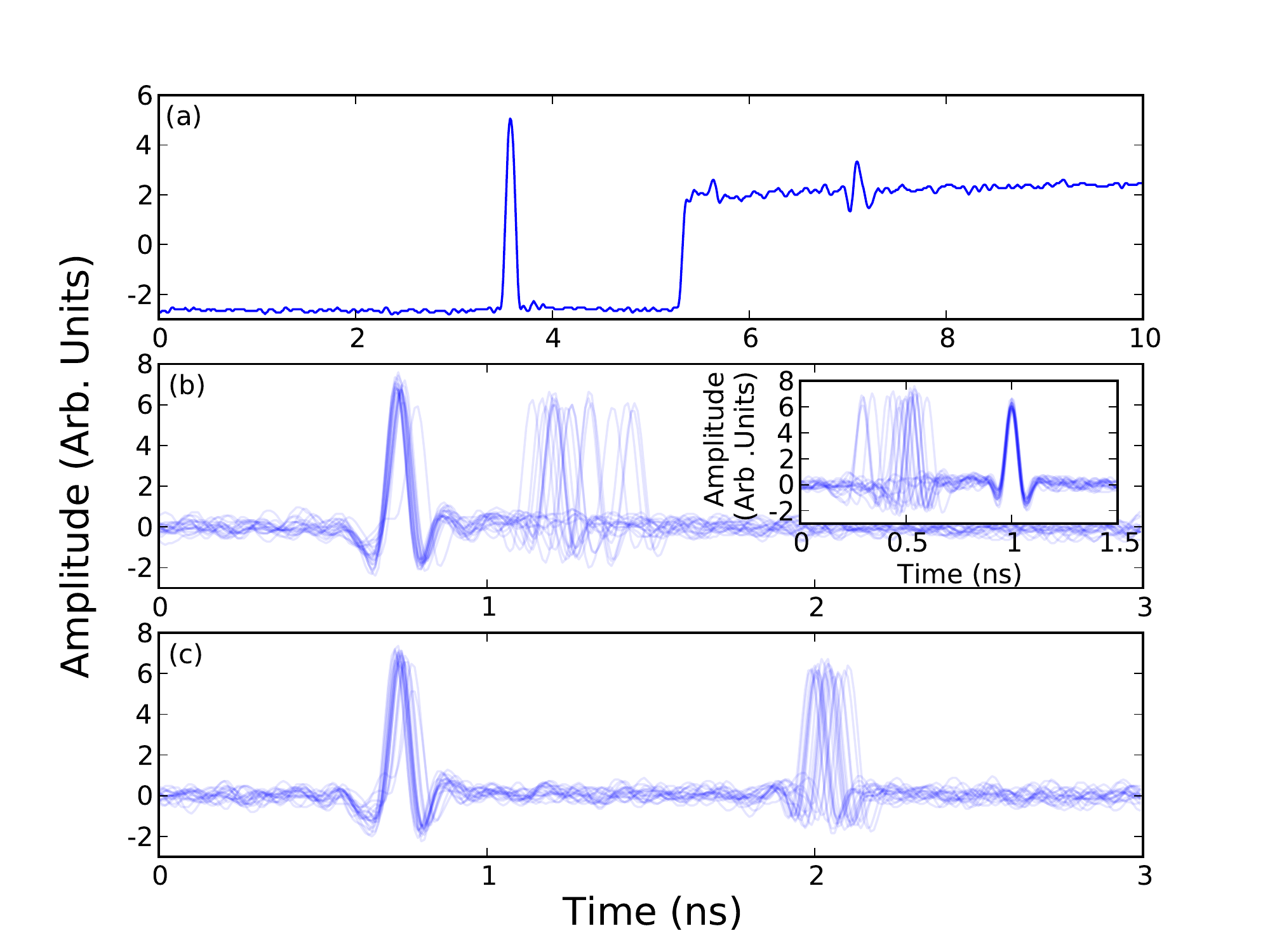}
\caption{Temporal traces of the perturbation and excitable responses (only when two responses are observed) for $I_\mathrm{sl} = 1.344$ mA,
$P_\mathrm{inj} = 6.3$ $\mu$W, $\Delta = 6.4$ GHz. (a): Shape of the perturbation for the maximum delay used, 1.81~ns. (b): Superposition of twenty excitable events for a perturbation delay of about 0.34~ns, triggered on the perturbations. Inset of (b): Same pulses as (b), triggered on the maximum of the second responses. (c): Superposition of twenty excitable events for a perturbation delay of about 1.2~ns, triggered on the perturbations.}\label{fig4}
\end{figure}
\begin{figure}[t]
\centering
\includegraphics[width=1.0 \columnwidth]{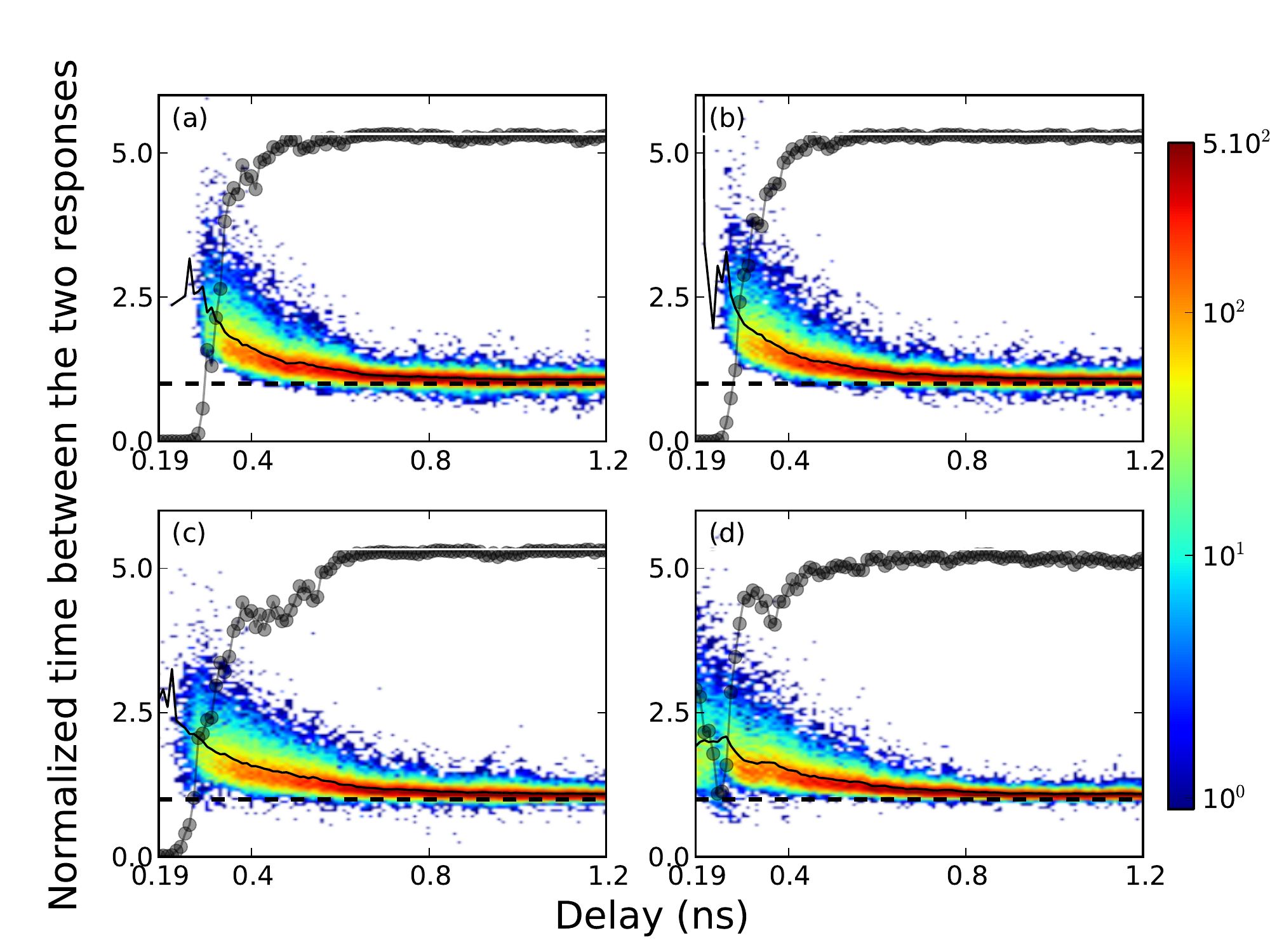}
\caption{Evolution of the histograms of the interval delay times between the two responses (only when 2 responses are obtained) in logarithmic color scale, normalized to the delay times of the perturbations. In each panel about 1600 events for each of 163 values of the delay are shown, for different values of parameters. Solid black line: mean of the distribution. Black dashed thick line: time normalized to the delay between perturbations. Grey line with circles: efficiency curve of the second pulse added as a guide, assuming the first has been triggered, the white lines indicates 100\% efficiency.
(a): $I_\mathrm{sl} = 1.344$ mA, $P_\mathrm{inj} = 6.3$  $\mu$W, $\Delta = 6.4$ GHz.
(b): $I_\mathrm{sl} = 1.338$ mA, $P_\mathrm{inj} = 8$    $\mu$W, $\Delta = 8.3$ GHz.
(c): $I_\mathrm{sl} = 1.329$ mA, $P_\mathrm{inj} = 10.2$ $\mu$W, $\Delta = 9.2$ GHz.
(d): $I_\mathrm{sl} = 1.321$ mA, $P_\mathrm{inj} = 12.7$ $\mu$W, $\Delta = 10.6$ GHz.}\label{fig5}
\end{figure}

The refractory time was determined by varying the time interval between the two perturbations. In Figs. \ref{fig4} and \ref{fig5} we analyse the role of this time interval for the nucleation of excitable pulses. In particular we found the minimum time interval which enables the nucleation of a second pulse, and, when the two excitable pulses exist, we compared their distance with the time interval of the two perturbations.

Most of the time we observed that one or two responses were induced by the two perturbations. Very rarely (in about 100 cases out of the $163\times1600=260800$ measurements in each panel of Fig. \ref{fig5}) more than two spikes have been observed in response to the two perturbations. In general those responses seemed not to be related to the perturbations so that, also due to a large distribution of the arrival times, we interpret them as noise induced pulses. However those events are so rare that we do not attempt to analyze them here and in Figs. \ref{fig4} and \ref{fig5} we report only the data associated with a double response.
	
Fig. \ref{fig4}(a) shows the shape of the electrical signal sent to the phase modulator for the maximum time interval we tested (1.81~ns). The amplitude of the perturbation is kept constant, close to the maximum amplitude used in Fig. \ref{fig3}(b). A little ringing is visible after the second edge, but it is not problematic for the experiment because of its small amplitude and very long delay. Fig. \ref{fig4}(b) shows a superposition of twenty excitable events for a time interval between the two perturbations of about 0.34~ns. Two excitable responses are observed in this temporal trace, corresponding to the same small intensity peak and phase rotation as before. Each one is triggered by a rising front in the perturbation profile, one related to the pulse and the other to the upward step. A large distribution of the delay time of the second pulse is observed.

The inset of Fig. \ref{fig4}(b) represents the same traces as in Fig. \ref{fig4}(b) triggered on the second pulses. Again, all the pulses are very similar, confirming their excitable character. In Fig. \ref{fig4}(c) the time interval between perturbations was increased to about 1.2~ns. The delay between the two responses is obviously increased, and we observed a narrow distribution on the arrivals time of the second pulse compared to Fig. \ref{fig4}(b).

Using the perturbation shown in Fig. \ref{fig4}(a) we perturbed the system repetitively, varying the time interval between the two edges.
We then measured the delay between the two excitable responses, when they exist, and reported this delay in Fig. \ref{fig5}.
One vertical cut of each panel of Fig. \ref{fig5} represents a histogram of the delays observed normalized to the time intervals between the two perturbations, realized over more than 1600 events. Fig. \ref{fig5} shows the evolution of such a histogram with the time interval between the two perturbations for four different combinations of injection strength and detuning. The evolution of the mean of the histogram is indicated by a solid black line.
Some typical responses of Fig. \ref{fig5}(b) were shown in Fig. \ref{fig4}(b,c). We add an efficiency curve (grey line with circle) for the second excitable pulse as a guide. Note that the absolute excitable threshold (for one pulse) is not the same for the four panels of Fig. \ref{fig5} since it depends on the parameters.

In general for lower time intervals between perturbations, the efficiency goes to zero abruptly meaning an impossibility to excite the system a second time. We identified here a minimal time interval beyond which it is no longer possible to create a second pulse: the refractory period.
Around the excitable threshold we observed the same behavior as in Fig. \ref{fig3}(b). The second excitable pulse could appear few nanoseconds after the first. A broad distribution of the time between the two pulses is visible.

We interpret these results as follow. If the second perturbation comes too early, the system does not have enough time to go back to the stable steady state. In the phase space picture of the circle (with the two fixed points) the system, when applying the second perturbation, has not yet reached the stable fixed point hence its distance from the unstable fixed point is larger and the excitability threshold is farther. We are then in the situation investigated in Fig. \ref{fig3}(b), the second pulse will take a longer delay to appear due to the time for the system to escape from the unstable equilibrium, or it could even not to be generated. In other words, the perturbation has to be higher if we want the same result found for the first one.

For larger time intervals between perturbations the efficiency of the second one (grey line with circles in Fig. \ref{fig5}) goes to 100\% (horizontal white line). The distribution becomes sharp and tends to a normalized time between the two pulses equal to one (horizontal grey dashed line). This indicates that the second pulse is created at a constant delay from the first, equal to the original distance between the two perturbations. A further analysis as developed in Fig. \ref{fig3}(b) could show that each pulse, in
this region, is also \lq\lq close"  to its perturbation due to the high amplitude of perturbation used. This could also be inferred by the narrow histograms in this region. Sometimes the normalized delay between responses is noticeably smaller than 1, which may seem surprising. However, it simply results from the fact that there is always some dispersion in the time at which the first pulse actually takes place, even if the perturbation is markedly beyond threshold. This results in a normalized delay between two responses smaller than unity when the first pulse is (due to noise) noticeably delayed with respect to the perturbation triggering it while the second is not.

Comparing the four panels of Fig. \ref{fig5} we can extract two interesting features.

First, the refractory period becomes shorter when increasing $P_{inj}$ and $\Delta$. We can understand this fact because for larger $\Delta$ excitable pulses are shorter, then the refractory period becomes shorter, too. It is then possible to trigger closer pulses.

Second, we notice a modification of the efficiency curve close to threshold with the increase of $P_{inj}$ and $\Delta$. Figs. \ref{fig5}(a,b) are similar, but we can see in Fig. \ref{fig5}(c) the appearance of a large hole, before reaching the 100\% efficiency. In Fig. \ref{fig5}(d) we are even not able to distinguish the refractory time because of a deep hole in the efficiency curve, followed by a little one. In fact the refractory period is composed of two times. A first time in which the system makes the $2\pi$ relative phase slip, and a second time in which the system relaxes to the stable fixed point through some oscillations due to the relaxation of the carrier. This feature is of course absent in the picture of the overdamped pendulum with forcing as described by Adler's equation since the inertial term which allows oscillations around the stable fixed point is absent.
\section{Theory and numerics}

\begin{figure}[b]
\centering
\includegraphics[width=1.0 \columnwidth]{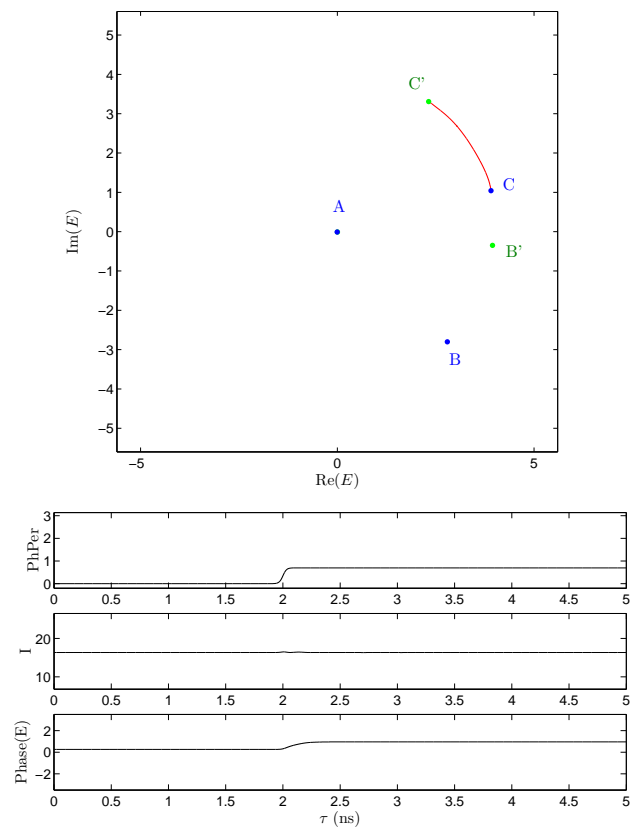}
\caption{Response of the slave laser to a step-up phase perturbation $\Delta\phi=40^\circ$ of the master laser electric field.
The upper panel shows the trajectory in the complex electric field plane. $A$, $B$ and $C$ are the fixed points before the perturbation and $B^\prime$ and $C^\prime$ those after. The lower panels show, from top to bottom, the time evolution of the phase of the master laser, and of the intensity and phase of the slave laser.}\label{figs40}
\end{figure}
The experimental findings were simulated using the following set of dynamical equations for an optically injected semiconductor laser
\begin{eqnarray}
\frac{dE}{dt} &=& \sigma\left[E_I+\left(1-i\alpha\right)DE-(1+i\theta)E\right]+\xi(t)\,,\\
\frac{dD}{dt} &=& \mu-\left(1+|E|^2\right)D\,,
\end{eqnarray}
where $E$ is the slowly varying envelope of the electric field, $D$ is the population variable proportional to the excess of carriers with respect to transparency, $\xi(t)$ is a Gaussian source of noise with $\langle\xi(t)\rangle=0$ and $\langle\xi^*(t)\xi(t')\rangle=\beta\delta(t-t')$, and $\alpha$ is the linewidth enhancement factor. Time is scaled to the carrier lifetime and $\sigma=\tau_c/\tau_p$ where $\tau_p$  is the photon lifetime. For the sake of simplicity and as we operate around a constant emission state, we neglect the fact that the spontaneous emission term $\beta$ may depend on the level of the population inversion $D$.

The three control parameters of the experiment here are denoted by $\theta$, $\mu$ and $E_I$.
$E_I$ is the dimensionless complex amplitude of the externally applied field, $\mu$ is the pump parameter of the slave laser proportional to the excess of injected current $I_{sl}$ with respect to the threshold $I_{th}$, and the cavity detuning $\theta$ is related to the experimental detuning $\Delta$ by
%
%

%
\begin{equation}
\theta=-\alpha+2\pi\Delta\tau_p=-\alpha+\frac{2\pi\Delta'}{\sigma}\,,\qquad \Delta'=\Delta\tau_c\,.
\end{equation}
Assuming $\tau_c=1$ ns, $\Delta'$ is just the detuning in GHz.
\begin{figure}[t]
\centering
\includegraphics[width=1.0 \columnwidth]{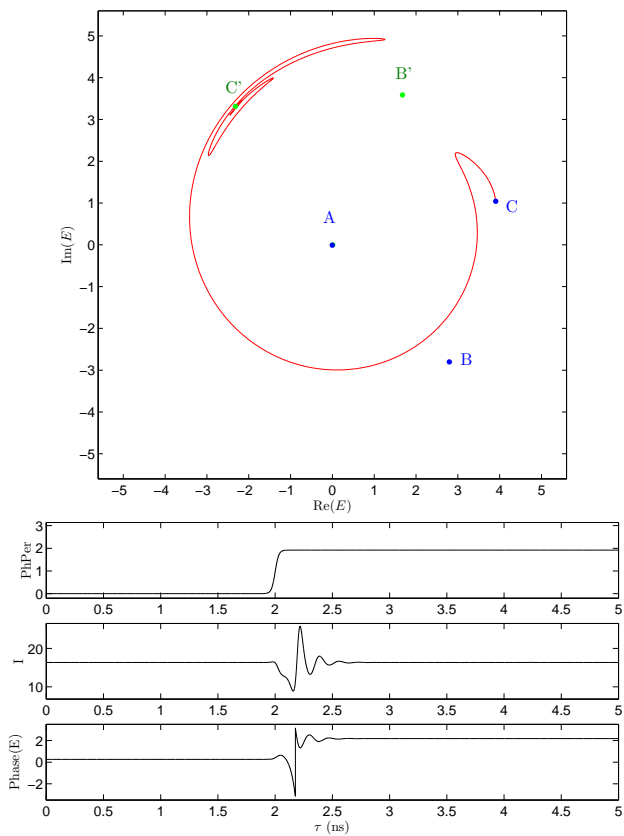}
\caption{Same as Fig. \ref{figs40} with $\Delta\phi=110^\circ$, which is above the excitability threshold. Here the trajectory from $C$ to $C^\prime$ follows a longer clockwise path, giving rise to an intensity pulse, as shown in the lower panels.}\label{figs110}
\end{figure}

In the numerical simulations we fixed $\alpha=4$, $\sigma=50$ (\textit{i.e.} $\tau_p=20$ ps if $\tau_c=1$ ns), $\mu=15$, and $\Delta'=4$ ($\theta=-3.4976$). With this choice of $\mu$ (which was not optimized but simply set to a high value in order to strongly damp the relaxation oscillations), $\alpha$ and $\theta$ the stationary state is bistable in the interval $0.238<|E_I|^2<755$.
We chose the value of input intensity $|E_I|^2=0.317$ in order to be close to the saddle-node transition corresponding to the left turning point
of the bistable input-output curve. Therefore, the system has three fixed points, that will be denoted as $A$, $B$ and $C$ in the following.
$A$ is an unstable focus very close to the origin in the complex plane, $B$ and $C$ are, respectively, a saddle and a stable node, close one to the other in the complex plane (they merge for $|E_I|^2=0.238$). Trajectories that leave $B$ along the unstable manifold end up in $C$ following approximately a circle.

\begin{figure}[t]
\centering
\includegraphics[angle=-90,width=1.0 \columnwidth]{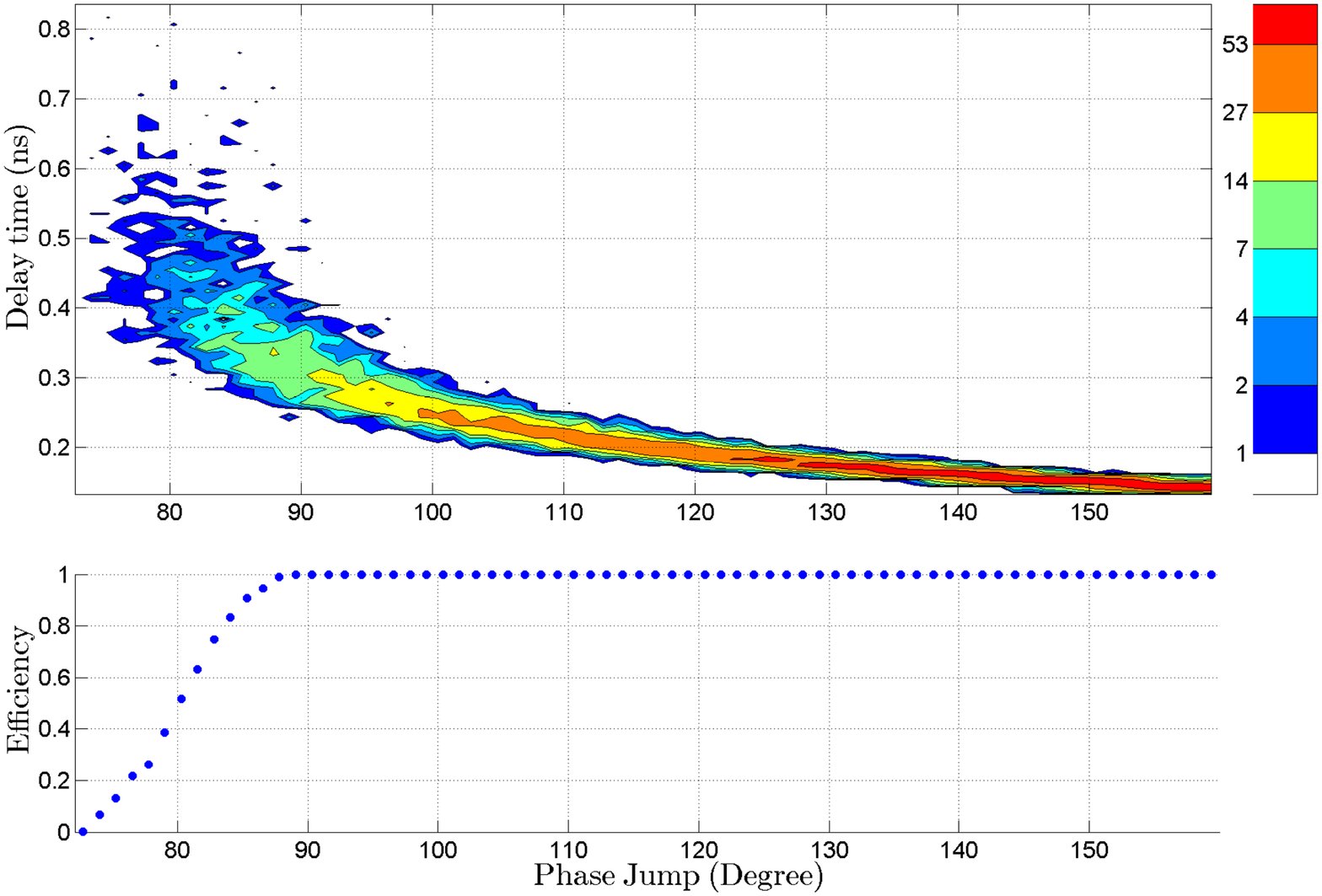}
\caption{Dependence on the perturbation amplitude of the time delay between the perturbation and the excitable response and of the efficiency, according to the numerical simulations. Same parameters as in Figs. \ref{figs40}, \ref{figs110}. Top: delay time histogram. Bottom: associated efficiency curve, defined as the number of excitable responses over the number of perturbations.}\label{efhistos}
\end{figure}

When we apply, as in the experiment, a step up phase perturbation $\Delta\phi$ to the input field, the three points rotate anticlockwise in the complex plane by the same angle. In Figs. \ref{figs40}, \ref{figs110}  we denote by $B^\prime$ and $C^\prime$ the rotated points ($A'$ is almost indistinguishable from $A$), which are the new fixed points of the system after the perturbation is applied. The response of the system depends on the magnitude of the phase step which is simulated as a sigmoid function with 50 ps rise time.

Fig. \ref{figs40} shows the trajectory in the complex plane when the applied perturbation $\Delta\phi=40^\circ$ is below the excitability threshold. Since the rotation is smaller than the angle between $B$ and $C$, the system simply moves from $C$ to $C^\prime$ along the shorter, anticlockwise path. The intensity remains essentially constant as the phase jumps to the new stationary value, as shown in the lower panels of the figure.

The dynamics is completely different when the applied perturbation is large enough to rotate the saddle beyond the old fixed point $C$, as in Fig. \ref{figs110}, where $\Delta\phi=110^\circ$. Now the presence of the new saddle $B^\prime$ in between $C$ and $C^\prime$ prevents the system from following the shorter path to $C^\prime$. Instead, the trajectory is captured by the unstable clockwise manifold of $B^\prime$ and the system reaches the new fixed point $C^\prime$ only after circling around the unstable focus. During the motion the distance from the origin reaches first a minimum, when the trajectory is closer to $B^\prime$, and then a maximum. Finally, the approach to $C^\prime$ is through damped relaxation oscillations. This explain the shape of the pulse intensity shown in the lower panel, which is very similar to the experimental traces, although there the first minimum is less deep. These differences can be attributed to the fact that the measured field consists of a superposition of the emitted laser field and of the reflection of the injected beam on the upper mirror of the VCSEL cavity.
\begin{figure}[t]
\centering
\includegraphics[width=1.0 \columnwidth]{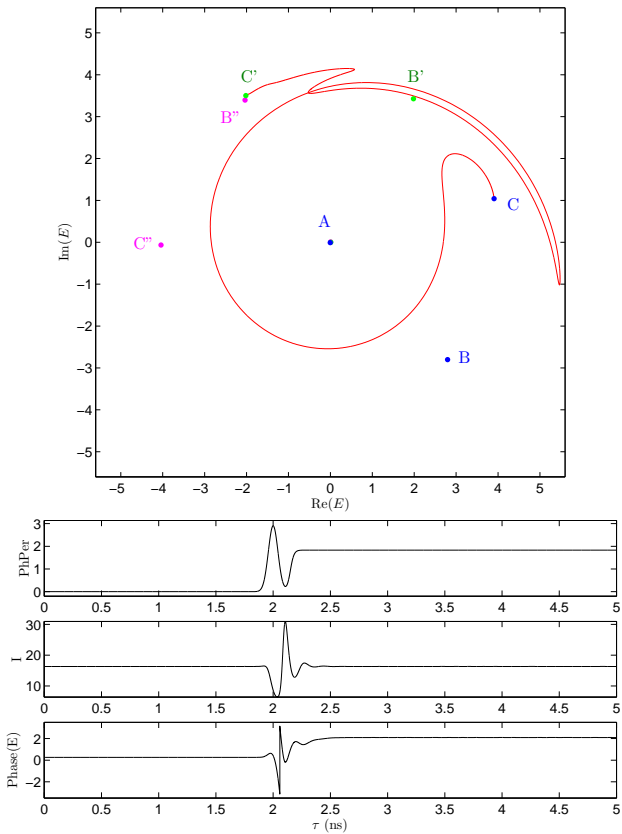}
\caption{Response of the slave laser to a double phase perturbation consisting of a narrow Gaussian impulse and a step-up perturbation. The delay of the two, which is 0.15 ns, is smaller than the refractory time and just one pulse is excited.  Same parameters as in Fig. \ref{figs40}. The labels $A$, $B$, $C$, $A'$ and $B'$ have the same meaning as in Fig. \ref{figs40}; $B''$ and $C''$ are the intermediate positions of the saddle and of the stable node when the pulse perturbation is applied.}\label{figsi015}
\end{figure}

In the simulations shown in Figs. \ref{figs40}, \ref{figs110} the noise terms were set to zero to better illustrate the deterministic dynamics.
Yet, in order to reproduce the experimental results a large number of simulations were performed with a realistic value of noise $\beta=100$. Fig. \ref{efhistos} is the counterpart of Fig. \ref{fig3}. It was built collecting the results of 9000 simulations with random phase perturbations uniformly distributed between $70^\circ$ and $160^\circ$. The behaviour of the histogram is very similar to the experimental one. Note that the delay shown here is the pure dynamical delay, whereas in Fig. \ref{fig3} it included that due to the detection apparatus. Hence, what matters is the increase of the delay time with respect to the asymptotic value found for large perturbation amplitudes, which is of the order of 0.5-1 ns in both the experimental and the numerical plots, although somewhat smaller in the latter. A small difference is found also in the excitability threshold, which is about $80^\circ$ in the numerical simulations and $65^\circ$ in the experiment.

\begin{figure}[t]
\centering
\includegraphics[width=1.0 \columnwidth]{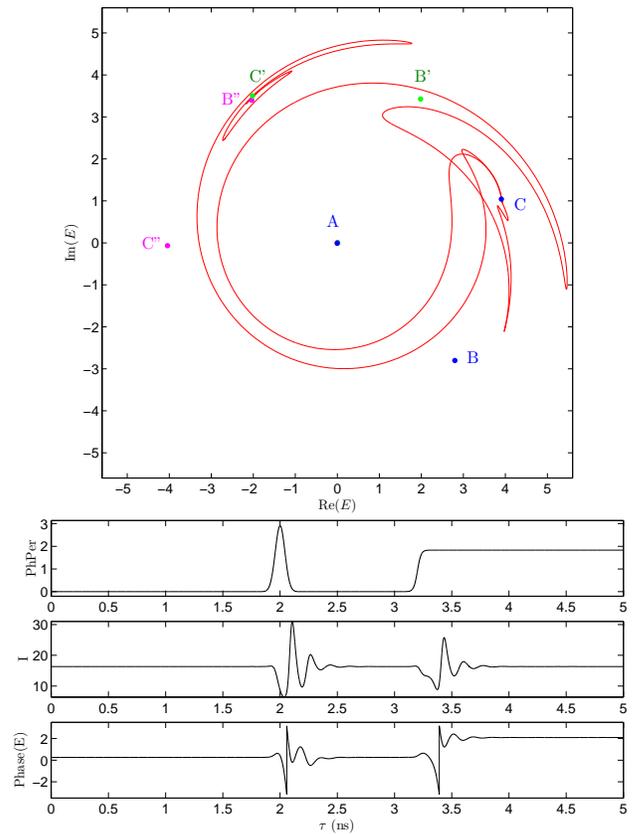}
\caption{Same as Fig. \ref{figsi015} with a delay of $1.2$ ns between the two perturbations, which is large enough to excite two pulses.}\label{figsi120}
\end{figure}

The experiment about the refractory period was simulated by applying to the phase of the master laser two perturbations. The first one was a short
Gaussian impulse with 100 ps FWHM and height $\Delta\phi=150^\circ$, while the second one was the same sigmoid function used before, with height $\Delta\phi=105^\circ$. In Figs. \ref{figsi015}, \ref{figsi120} the points $B''$ and $C''$ indicate the temporary positions of the saddle and of the node when the Gaussian impulse is applied. 
In excellent agreement with the experiment, we found that the refractory time is about 0.2 ns.
Fig. \ref{figsi015} shows the dynamics for a smaller delay between the two perturbations, 0.15 ns. The trajectory in the complex plane at the beginning is similar to that of Fig. \ref{figs110}: after being captured by the unstable manifold of the saddle $B''$ the system moves clockwise toward the node $C''$. Yet, because of the short duration of the first pulse, $C''$ quickly moves to the original position $C$ and the trajectory follows it. In this way an intensity pulse is generated. However, the node moves to the final position $C'$ before the system can actually reach it, because the delay between the two perturbations is very short. During the rotation of the fixed points due to the step-up perturbation the trajectory remains trapped between $B'$ and $C'$ and finally it moves to $C'$ following the short anticlockwise path, \textit{i.e.} without generating a pulse.

In Fig. \ref{figsi120} the delay between the perturbation is 1.2 ns, much larger than the refractory time. The dynamics of the first pulse is similar to that of Fig. \ref{figsi015}, but in this case after the pulse the system has all the time to return to the initial position $C$ of the node. Hence, when the step-up perturbation is applied, a second pulse is generated, similarly to what is shown in Fig. \ref{figs110}.

\begin{figure}[t]
\centering
\includegraphics[angle=0,width=1.0 \columnwidth]{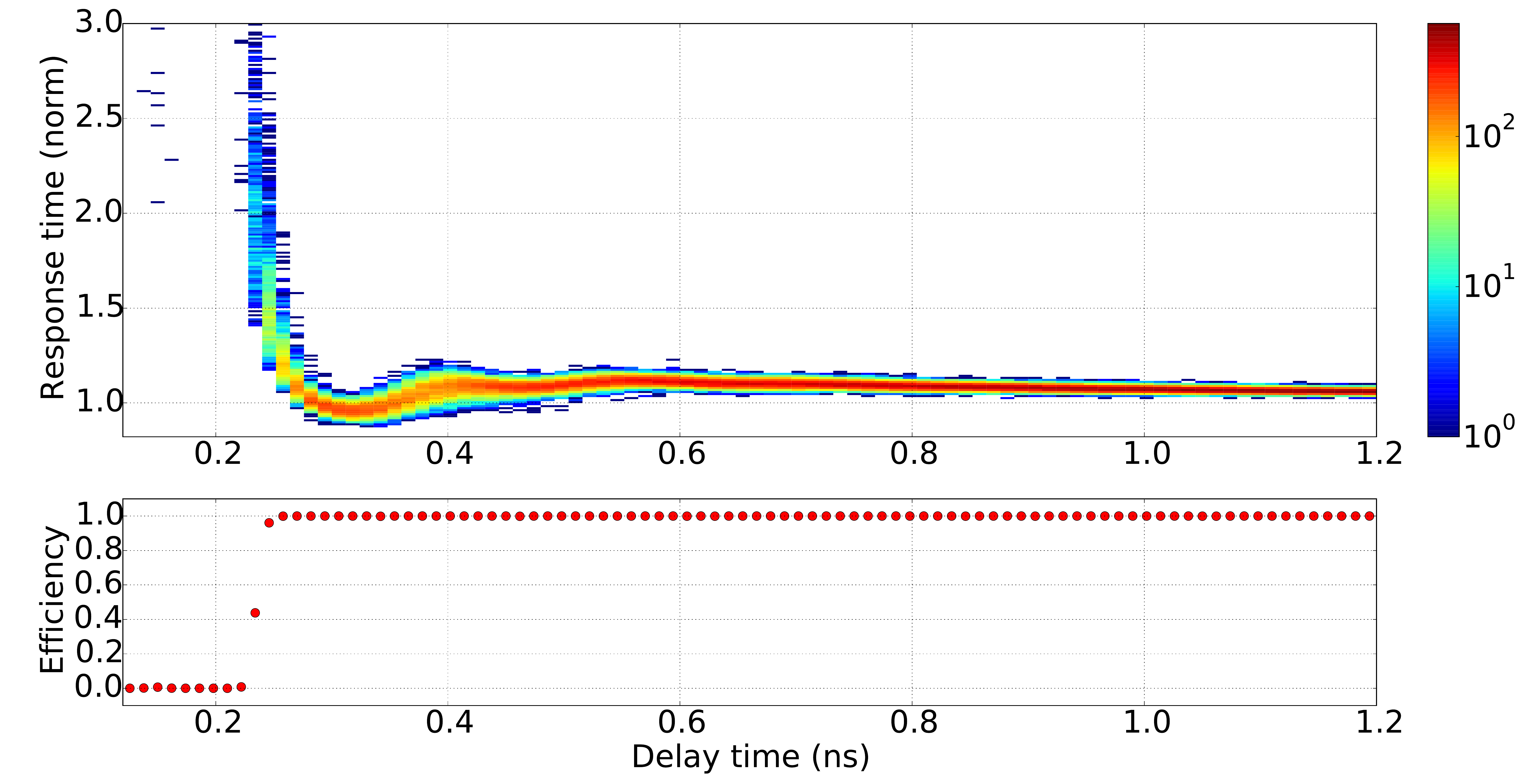}
\caption{(a) Delay between the first and second excitable pulses and (b) efficiency of the generation of the second pulse as a function of the delay time of the perturbations. Same parameters as in Figs. \ref{figsi015}, \ref{figsi120}. The delay in the vertical axis is normalized to the delay of the perturbations, so that the dashed horizontal line represents the situation where the delay of the excitable pulses equals that of the perturbations.}\label{efhistosi}
\end{figure}

By integrating the dynamical equations with noise amplitude $\beta=100$  we were able to compare the delay of the excitable pulses with respect to the temporal separation between the perturbations. The results are shown in Fig. \ref{efhistosi}, which is analogous to Fig. \ref{fig5}. The efficiency plot clearly shows that the probability of having a second excitable pulse falls to zero when the delay of the perturbations is smaller than 0.22~ns. The histogram also shows that, as in the experiment, approaching the refractory time the temporal separation between the excitable pulses can be more than twice the temporal separation between the perturbations. As in the experiment, the width of the histogram of the arrival time for the second pulse also increases when the separation between the perturbation diminishes and gets closer to the refractory time.

Finally, by considering the dynamics of the system shown in Fig. \ref{figsi120} we can propose an explanation for the non-monotonicity of the most probable value of delay between responses in Fig. \ref{efhistosi}a) and in the efficiency curves in panels (c,d) of \ref{fig5}. When the delay between the perturbations approaches the refractory time, the system is still relaxing towards the node $C$ via usual semiconductor laser relaxation oscillations when the step-perturbation is applied. Hence, for slightly different time separations between the perturbations the system will be in different positions around $C$, some of them more favorable, and some less for the generation of the second pulse. This is also the reason for the very exceptional events taking place for a delay between perturbations of about 0.1~ns which can be observed as isolated dots on Fig. \ref{efhistosi}a). Of course, the description of these effects is not possible in the simple vision of a pure phase oscillator and we keep a detailed analysis of this point (which becomes much more pronounced when relaxation oscillations are less damped) for future work.

\section{Conclusions}
In conclusion we have analyzed experimentally the response of a semiconductor laser with optical injection in the excitable regime to one or two successive perturbations applied in the phase of the driving field. Our experimental data show that when the perturbation is hardly able to nucleate an excitable pulse the dispersion of the response time increases but, in agreement with the excitable nature of the system, the response themselves remain perfectly identical to each other. In addition, when two perturbations are applied in short sequence, the second may not be able to trigger a response, showing the existence of a refractory time in this system. In analogy with what happens for a single perturbation, hardly successful stimulations will be followed by responses which are strongly dispersed in time yet all identical to each other, illustrating the nature of the refractory period. These observations may provide elements of interpretation for the repulsion between phase bits observed when submitting a laser with injected signal in the excitable regime to delayed optical feedback \cite{garbin2015topological}.

The experimental results have been interpreted by means of numerical simulations of a set of dynamical equations for an optically injected semiconductor laser, realized in parameter ranges compatible with the experimental ones \textit{i.e.} in a regime of strongly damped relaxation oscillations.

\bibliography{rp}

\end{document}